\documentclass[aps,prb,twocolumn,groupedaddress,floats,showpacs]{revtex4-1}
\usepackage{latexsym}
\usepackage{dcolumn}
\usepackage[dvips]{graphicx}
\usepackage{amssymb}
\usepackage{graphics}
\usepackage{amsmath}
\usepackage{epsf}

\newcommand{\vk}{{\bf k}}
\newcommand{\vq}{{\bf q}}

\newcommand{\beq}    {\begin{equation}}
\newcommand{\enq}    {\end{equation}}

\begin{document}
\title{Non-monotonic temperature dependent transport in 
graphene grown by Chemical Vapor Deposition}
\author{J. Heo$^1$, H. J. Chung$^1$, Sung-Hoon Lee$^1$, H. Yang$^1$,
  D. H. Seo$^2$, J. K. Shin$^1$, U-In Chung$^1$}
\author{S. Seo$^{1}$}
\email{sunaeseo@samsung.com}
\author{E. H. Hwang$^3$}
\author{S. Das Sarma$^3$} 
\affiliation{$^1$Semiconductor Devices Lab, Samsung Advanced Institute of
  Tech., Giheung-Gu, Yongin-Si, Gyeonggi-Do 449-712, Korea}
\affiliation{$^2$Solid State and Photonics Lab, Stanford University,
  Stanford, CA  94305, USA} 
\affiliation{$^3$Condensed Matter Theory Center, Department of 
        Physics, University of Maryland, College Park, MD 20742-4111}

\begin{abstract}
Carrier density and temperature-dependent resistivity of graphene grown by chemical vapor
deposition (CVD) is investigated. We observe in low mobility CVD
graphene device a generic insulating behavior at low temperatures, and
eventually a
metallic behavior at high temperatures, manifesting a non-monotonic 
temperature dependent resistivity.  This feature is strongly
affected by carrier density modulation with the low-density samples
exhibiting insulating-like temperature dependence upto higher
temperatures than the corresponding high-density samples. To explain 
the temperature and density dependence of the resistivity, we
introduce thermal activation of charge 
carriers in electron-hole puddles induced by randomly distributed
charged impurities. Our observed temperature evolution of resistivity is
then understood from the competition among thermal activation of
charge carriers, temperature-dependent screening, and phonon scattering
effects. Our experimental results are in good agreement with recent
theories of graphene transport.
\end{abstract}
\maketitle

\section{introduction}

Graphene, a two dimensional honeycomb lattice of carbon atoms composed
of two identical interpenetrating triangular sub-lattices, has
attracted research interest because of its novel linear dispersion
relation resulting in unique electronic properties such as room temperature
intrinsic carrier
mobility\cite{dassarma:2011,novoselov:2004,geim:2007,bolotin:2008}. Exploiting such
superior electronic 
properties combined with recently developed large scale growth of
graphene by Chemical Vapor Deposition
(CVD)\cite{reina:2009,kim:2009,li:2009}, has led to field 
effect devices integrated on a wafer scale showing promise for future
electronic applications\cite{lee:2010}. However, field effect performance of
exfoliated graphene is adversely affected by carrier scattering by
charged impurities in the surrounding media\cite{chen:2008,adam:2007}, 
surface phonons\cite{chen:2008a,zou:2010,fratini:2008} on gate
dielectric, and graphene acoustic
phonon\cite{hwang:2008,chen:2008a}. Although graphene grown by CVD 
shows outstanding uniformity and mobility on a wafer scale, field
effect transport measurement performed after the transfer process, may
be strongly affected by undesirable impurity scattering\cite{lee:2011}.
Therefore, understanding transport on CVD-grown graphene influenced by
impurity scattering is an important issue for funamental physics and
for further improvement of
device performance.

In this work we are presenting an experimental study of monolayer
graphene transport properties, with the emphasis on understanding the
temperature dependent resistivity, complemented by a detailed
comparison with theoretical calculations of graphene resistivity
limited by charged impurity and phonon scattering. Such a combined
experimental and theoretical study has not been undertaken for CVD
graphene. Our work is closest in spirit to the early work by Tan {\it
  et al.}\cite{tan:2007} in exfoliated graphene where, however, only
the low-temperature density-dependent transport properties of graphene
were studied in a combined experimental-theoretical investigation to
establish the dominant role of charged impurity scattering in
determining the low-temperature graphene electrical
conductivity. Early work on graphene transport properties
\cite{novoselov:2004,geim:2007} emphasized the weak temperature
dependence of graphene resistivity upto the room temperature,
presumably because of the weakness of the basic electron-phonon
coupling in graphene. Such a weak temperature dependent conductivity is
unheard of in electronic materials where phonons invariably dominate
the room-temperature resistivity. In graphene, the very weak phonon
effects on the electrical conductivity have only recently been
experimentally observed \cite{efetov:2010} verifying earlier
theoretical predictions \cite{hwang:2008}. 
Although there have been several earlier experimental works on the
temperature dependence of transport properties in {\it exfoliated}
graphene, our current work is the only work in the literature, to the
best of our knowledge, which reports on the measured temperature
dependence of graphene resistivity in {\it CVD} graphene. 
Our work is also the only one in the literature combining experiment
with detailed theory to shed light on the physical processes
controlling the temperature dependence of graphene resistivity. One
of our most important findings is that the low carrier density (and
low temperature) temperature-dependence in graphene is dominated to a
large extent by carrier activation across the potential fluctuations
created by random charged impurities in the environment (which also
lead to the inhomogeneous electron-hole puddle formation responsible
for the graphene minimum conductivity phenomenon). We compare our
experimental results with a recent theory \cite{li:2011} on the puddle
induced temperature dependence in graphene transport
properties. obtaining excellent agreement. Such a comparison between
theory and experiment for graphene transport at low temperature (and
densities) where puddles dominate transport properties is completely
new in the literature.

One other important indirect consequence of our work is the conclusion
that graphene resistivity seems to manifest similar qualitative
temperature and density dependence for {\it similar mobility samples}
independent of how it is made, i.e. our CVD-grown samples exhibit a
temperature dependent resistivity qualitatively similar to exfoliation
graphene samples of similar mobility. This establishes beyond any
doubt that the factor determining graphene transport properties (both
density and temperature dependence) is the effective
``zero-temperature'' (i.e. very low-temperature) mobility which itself
is determined by the charged impurity distribution in the
environment. Thus, improving graphene mobility for electronic
applications depends entirely on making graphene purer by removing the
impurity content from the environment. Having CVD graphene helps only
if it has low impurity content, which, according to our work, is no
longer the case when CVD graphene is transferred.

We investigate temperature ($T$) dependent resistivity, $\rho(T)$, of
graphene grown by CVD at various carrier densities ($n_g$). In low
mobility samples, a strong insulating behavior at low temperature and
a metallic behavior at high temperature is observed, resulting in
non-monotonicity. (We define “insulating”/“metallic” as $d\rho/dT$ being
negative/positive in the sample -- this is purely an operational
definition which is used extensively in the literature: all our
samples are metallic in reality since a mobility can be defined.) 
This feature persists up to
relatively high carrier density. On the other hand, high mobility
samples show overall metallic behavior from fairly low to high carrier
density. To understand this peculiar transport property of CVD
graphene, we introduce the concept of thermal activation of charge
carriers in electron-hole puddles generated by charged impurity\cite{hwang:2010}
which strongly affects the graphene mobility. Anomalous $T$ and $n_g$
dependence of transport is well described theoretically by the
competition among thermal activation, screened impurity scattering,
and phonon scattering. Thus, puddles play a very important role in
controlling the temperature dependent CVD graphene transport.  
We compare our data with a recently developed detailed theory
\cite{li:2011}. 

Temperature dependence of carrier transport properties,
e.g. resistivity, in electronic materials is often attributed to be
arising from carrier-phonon interaction since phonons are thermally
excited quantized-vibrations of the underlying lattice
\cite{dassarma:2011}.  For example, 
the temperature dependence of electrical resistivity in normal metals
like Al or Cu is essentially entirely controlled by the
electron-phonon interaction strength in the metal.  Such `metallic'
temperature dependence always involves an increasing resistivity with
increasing temperature, i.e. $d \rho/dT> 0$, and often, the definition
of a metal is taken to be synonymous with the $d\rho/dT >0$ behavior
(or very weak temperature dependence of resistivity)  with a finite
(and presumably, not too large) zero-temperature extrapolated
resistivity.   An insulator, on the other hand, has infinite (zero)
resistivity (conductivity) at $T=0$, and increasing $T$ typically
increases its conductivity since activated or variable-range hopping
transport become allowed at finite temperature.  An insulator or a
localized system thus has $d\rho/dT<0$  with an extrapolated
zero-temperature resistivity which is infinite (or very large).  Thus,
$d\rho/dT < 0$ is often taken to be an operational definition of an
insulator.  This operational definition of an effective 'metal'
($d\rho/dT >0$) versus an effective 'insulator' ($d\rho/dT<0$) is general
and powerful, and has long guided experimental and theoretical
research in electronic materials (metals, semiconductors, doped
materials). 
However, we emphasize that 'metallic' and 'insulating' used in our paper are
terminologies for the temperature dependence of resistivity. 
($d\rho/dT$ being negative/positive does not necessarily indicate a $T=0$ true
insulator/metal.)

Given the above very general consideration, it is therefore always
very interesting and intriguing when an otherwise well-defined
metallic system exhibits insulating (i.e. $d\rho/dT<0$) temperature
dependence.  It turns out that graphene is indeed such a rather
unusual electronic material which often manifests $d\rho/dT<0$ in some
carrier density and temperature range while exhibiting the expected
'metallic' behavior $d\rho/dT>0$ in other regimes of density and
temperature\cite{dassarma:2011,chen:2008,hwang:2008,chen:2008a,zou:2010}.  In
fact, for some samples (and at some specific values 
of carrier density) even non-monotonic temperature dependence with
resistivity decreasing with $T$ first and then increasing with $T$ is
observed.  In the current paper, we present a detailed experimental
study of the temperature dependence of graphene transport properties
comparing it with a quantitative theoretical analysis to show how and
why such interesting temperature dependence could arise.  The
non-monotonicity in the graphene resistivity indicates competing
physical processes, and indeed, there are several independent
contributions to graphene transport which all depend non-trivially on
density and temperature.  These physical processes combine to give
rise to the observed temperature dependence as we show in the current
work.  We emphasize that although some aspects of this temperature
dependence were earlier studied experimentally in the literature for
exfoliated graphene on substrates and for suspended graphene
\cite{chen:2008a,zou:2010,bolotin:2008a,zhu:2009}, ours is 
the first such temperature-dependent transport study for graphene
grown by the CVD technique. 
Ours is also the first experimental study where the measurements are
carefully compared with a detailed theoty.
 
The competing physical mechanisms contributing to the temperature
dependent graphene transport are phonons (which are known to be weak
in graphene), screening, Fermi surface effects, and carrier activation
across potential fluctuations associated with the density
inhomogeneities or puddles in graphene which dominate its low-density
properties.  Of these four mechanisms, the first two (phonons and
screening) always produce ‘metallic’ behavior whereas the last two
(Fermi surface and activation) always produce 'insulating' behavior.
In general, both insulating effects (i.e. Fermi surface effects and
activation across puddles) are stronger at low carrier densities
whereas the two 'metallic' mechanisms are stronger at higher carrier
densities.  Thus, we expect graphene to preferentially exhibit
metallic (insulating) behavior at higher (lower) carrier densities.
Of course, at high enough temperature, phonons will eventually
dominate, and all graphene samples should eventually exhibit metallic
transport properties (i.e. $d\rho/dT>0$), and therefore, at some
densities we expect to see non-monotonicity in the graphene resistivity
as a function of temperature.  All these expectations are beautifully
manifested in our experiments and we find reasonable quantitative
agreement with theory.  It is interesting and nontrivial, but
eventually quantitatively understandable, that graphene shows a
complex interplay of effective insulating and metallic behaviors
superposed on each other as a function of density with the insulating
(metallic)  behavior dominating low (high) densities without any real
localization or gapped insulator physics being present in the system.
Ours is the first complete study of this phenomenon in CVD graphene
with direct comparison between experiment and theory. 

The paper is organized as follows. In Sec. II we describe the sample
preparation and experimental results for resistivity. In Sec. III we
provide the theoretical analysis of measured non-monotonic
resistivity. We conclude in Sec. IV with discussion. 

\section{experiments}

Graphene devices in this study are synthesized by Inductively Coupled
Plasma enhanced Chemical Vapor Deposition (ICP-CVD) on Cu
substrate. During growth process, the substrate is heated to 650$^\circ$C
within 10 min under $\sim 10^{-7}$ torr, then treated with H$_2$ plasma. After
purging with Ar for a couple of minutes, C$_2$H$_2$ is added (C$_2$H$_2$: Ar =1:
40) for graphene growth at the same temperature. For the graphene
transfer, Graphene/Metal/SiO$_2$/Si substrate was spin-coated with PMMA
(Aldrich, 950 A4) and attached a pressure sensitive adhesive
ultraviolet-tape. Peeling the tape against Si wafer physically
separates Tape/PMMA/Graphene/Metal layer due to poor adhesion of metal
film and SiO$_2$. After etching of underlying Ni/Cu by soaking in FeCl$_3$
and cleaning in water, Tape/PMMA/Graphene layer was pressed onto the
SiO$_2$(300nm)/Si substrate with pre-patterned electrodes (Cr/Au of
10/50nm). The successive removal of tape and PMMA in methanol and
acetone, respectively, leaves only graphene layer over pre-patterned
marks. Graphene is etched to rectangular shape by O$_2$ plasma as shown
in the inset of Fig. 1(a). The distance between source and drain
electrodes is 8$\mu$m and the width of channel is 7$\mu$m. The T-dependent
resistivity is recorded using 4-probe measurement technique for $4K < T
< 300K$ under $\sim$10-5 torr after 24 hour annealing at 150 $^{\circ}$C. This
annealing process before measurement gets rid of hole-doping, and
moves the Dirac Point to near zero gate voltage ($V_g$).  

Single layer graphene was confirmed by Raman spectroscopy\cite{ferrari:2006} with
single Lorentzian fit of 2D peak at 2700 cm$^{-1}$ in Fig. 1(a)  and
optical transmittance of 97.7\% at 550nm \cite{nair:2008} in
Fig. 1(b). The optical image and Raman mapping from our previous work
\cite{lee:2011} verified that transferred CVD-grown graphene has no
sign of wrinkles or bi, tri-layers of graphene. According to
Matthiessen’s rule, we assume the conductivity is the sum of
contributions from long-range scatterers and short range scatterers
or residual resistivity. Hole mobility is then extracted using the
following formula,  
\begin{equation}
\sigma(V_g)^{-1}=-\left [ \mu_h C_g(V_g-V_{\rm Dirac}) \right ]^{-1} +
\sigma_{res,h}^{-1} \;\;\; V_g < V_{\rm Dirac},
\end{equation}
where $\sigma(V_g)$ is conductivity, $\mu_h$ is hole mobility, $C_g$
is capacitance of gate 
dielectric, $V_{\rm drag}$ is the gate voltage at Dirac point, and $\sigma_{res,h}$ is
residual conductivity. The 
measured hole mobility in a specific graphene sample in the same
6-inch wafer has a large variation by an order of magnitude. We
speculate this large variation is coming from non-uniform transfer
procedure which introduces local charge impurities differently. Two
groups of devices with mobility of $1,000 \sim 3,000$ cm$^2$/Vs (A) and
$10,000 \sim 13,000$ cm$^2$/Vs (B) respectively were selected to understand the
relationship between $\rho(T)$ at various $n_g$ and the level of disorder
(“high” in A, “low” in B)\cite{chen:2008,adam:2007}. Fig. 1(c) and (d)
show $V_g$-dependent 
resistivity [$\rho(V_g)$] of A and B devices, respectively. No marked
difference is observed in Raman spectroscopy of these two disparate
groups despite the order of magnitude variation in the mobility
supporting introduction of non-uniform charge impurities during
transfer. In addition, we observe a relatively weak D-peak, which
indicates minimum level of structural defects in our samples.

\begin{figure}
\includegraphics[width=\columnwidth]{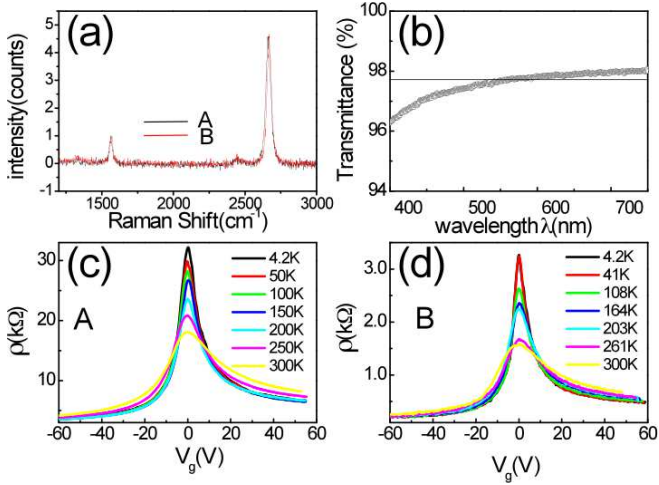}
\caption{
(color online) (a) Raman spectroscopy on a device, inset: Optical
  microscope image of a device. (b) Light transmittance through CVD
  grown single layer graphene/Glass. (c) and (d) resistivity vs. gate
  voltage at various temperature for the device group of  A and B,
  respectively.  
}
\end{figure}

T-dependence of resistivity [$\rho(T)$] with reference to $\rho(4.2K)$
at carrier densities near charge neutrality point is plotted in
Fig. 2(a) and (b) for A and B respectively. $\Delta V_g$ is equal to
$V_g-V_{\rm Dirac}$. The
characteristic observed feature of carrier transport in our low
mobility sample A (Fig. 2(a)) is a strong insulating behavior (i.e.,
$d\rho(T)/dT < 0$) at low temperatures ($T < 200K$) for all densities (up to
$7.2 \times 10^{11}cm^{-2}$). For $\Delta V_g > 4V$ ($n_g > 2.9 \times
10^{11}cm^{-2}$) a non-monotonicity in 
temperature dependent resistivity develops (i.e. as T increases, $\rho$
initially decreases and then increases forming a local minimum). The
observed insulating behavior and non-monotonicity at high density in
low mobility samples is not observed in high mobility samples B
(Fig. 2(b)), where the measured resistivity shows a metallic
temperature dependence at higher densities even though the
experimental conductivity manifests an insulating behavior at very low
carrier density (up to $n_g \sim 2.0 \times 10^{11}cm^{-2}$). Interestingly, negative
$d\rho(T)/dT$ at Dirac point is $\sim 40\%$ for both samples independent of sample
mobility, which also has been reported in high mobility suspended
graphene\cite{bolotin:2008a,du:2008}. Recently, ballistic transport
was invoked in order to 
explain the strong T-dependent $\rho(T)$ at Dirac point as well as low $n_g$
regime\cite{muller:2009}. However, this approach is not applicable in our low
mobility devices which are obviously in the diffusive regime. The
temperature dependence in our diffusive samples is presumably
connected with extrinsic effects such as charged impurity scattering,
which is considered the dominant scattering process for transport at
low $n_g$. Note that only metallic behavior has been seen in graphene
samples at high densities. 

Another important aspect of temperature dependent resistivity is that
the initial insulating behavior at Dirac point is strongly affected by
carrier density modulation. Starting from the charge neutrality point,
the insulating behavior becomes non-monotonic near $\Delta V_g \sim 4V$ ($n_g =
2.9 \times 10^{11}cm^{-2}$) for sample A due to the coexistence of low T insulating
and high T metallic behavior (Fig. 2(a)), but for sample B, a complete
transition to metallic behavior occurs already around $n_g = 2 \times
10^{11}cm^{-2}$ (Fig. 2(b)). 

\begin{figure}
\includegraphics[width=\columnwidth]{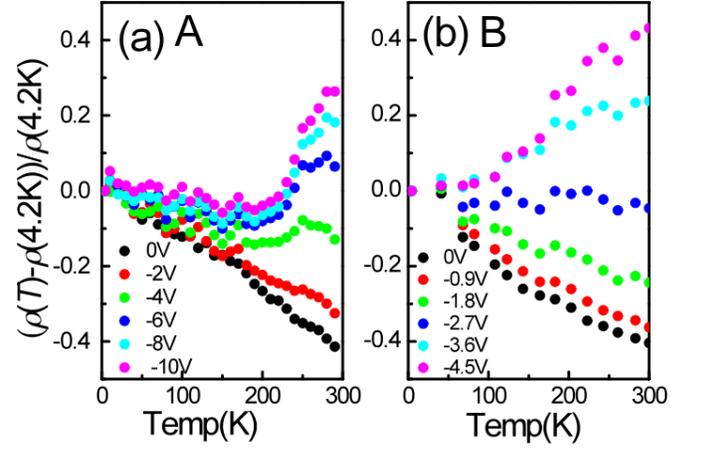}
\caption{(color online)
$[\rho(T)- \rho(4.2K)]/ \rho(4.2K)$ at varied $\Delta V_g$. (a) device
  A, (b) device B.}
\end{figure}

\section{theoretical analysis}

To understand T and $n_g$ dependence of different mobility devices, we
consider the quantitative theory for carrier transport in the presence
of electron-hole puddles due to the spatially fluctuating
inhomogeneous potential induced by random charged impurities. The
formation of electron-hole puddles has been proposed
theoretically\cite{hwang:2007} 
and observed in scanning tunneling spectroscopy\cite{zhang:2009} and scanning
single electron transistor microscopy experiments\cite{martin:2008}. The presence of
electron-hole puddles was recently introduced to explain the
insulating behavior in metallic bilayer graphene\cite{hwang:2010}.  It is crucial
to include the role of electron-hole puddles to explain the measured
insulating behavior in our monolayer graphene samples. We incorporate
the puddles into the potential fluctuations which is assumed to be
described by a statistical distribution function of a Gaussian form
with a standard deviation $s$.\cite{hwang:2010}  Thus, we calculate
theoretically the 
carrier transport for our graphene samples by taking into account the
highly inhomogeneous density and potential landscape.  We also include
the scattering by screened charged impurities, longitudinal acoustic
phonon scattering and surface optical phonon scattering. Then the
total scattering time is calculated by  
\begin{equation}
\frac{1}{\tau_{tot}}=\frac{1}{\tau_{imp}} + \frac{1}{\tau_{ap}} +
\frac{1}{\tau_{sp}}
\end{equation}                   
where $\tau_{imp}$, $\tau_{ap}$, and $\tau_{sp}$ are the transport
scattering times of screened charged impurity\cite{hwang:2009}, longitudinal
acoustic phonon\cite{hwang:2008}, and surface optical phonon\cite{fratini:2008},
respectively. 
The scattering time due to charged impurity $\tau_{imp}$ is given by
\begin{eqnarray}
\frac{1}{\tau_{imp}(\varepsilon)} = \frac{2\pi n_i}{\hbar} \int \frac{d^2
  k'}{(2\pi)^2} \left |\frac{v_i(q)}{\epsilon(q)} \right |^2
g(\theta_{\vk  \vk'}) \nonumber \\ 
\times [1-\cos\theta_{\vk\vk'}]
\delta\left (\varepsilon_{\vk} - \varepsilon_{\vk'} \right ),
\label{tau}
\end{eqnarray}
where $n_i$ is the charged impurity density, $\varepsilon_{\vk}=\hbar v_F
\vk$ is the carrier energy with the Fermi velocity $v_F$ and 2D wave vector $\vk$,
$g(\theta_{\vk,\vk'})=[1+\cos(\theta)]/2$ is a wave function form
factor associated with the chiral matrix of graphene,
$\theta_{\vk \vk'}$ is the scattering angle between $\vk$ and
$\vk'$, $v_i(q) = 2\pi e^2/\kappa q$ is
the 2D Coulomb potential, and $\epsilon(q)$ is the finite temperature
dielectric screening function.
The scattering time due to acoustic phonon mode $\tau_{ac}$
is given by
\begin{equation}
\frac{1}{\tau_{ac}(\varepsilon)} = \sum_{\vk'}(1-\cos\theta_{\vk \vk'}) W_{\vk
  \vk'}\frac{1 - f(\varepsilon')}{1-f(\varepsilon)},
\end{equation}
where $f(\varepsilon)=[e^{(\varepsilon-\mu)/k_BT} + 1]^{-1} $ is the Fermi
distribution function with the chemical potential $\mu$,
and $W_{\vk \vk'}$ is the transition probability from the state with 
momentum $\vk$ to $\vk'$ state and given by
\begin{eqnarray}
W_{\vk \vk'} = \frac{2\pi}{\hbar}\sum_{\vq}|C(\vq)|^2 &\times&
  [N_q \delta(\varepsilon-\varepsilon'+\omega_{\vq}) \nonumber \\ 
& + & (N_q + 1) 
\delta(\varepsilon-\varepsilon'-\omega_{\vq})  ],
\label{wq}
\end{eqnarray}
where $C(\vq)$ is the matrix element for scattering by acoustic phonon,
$\omega_{\vq}=v_{ph} \vq$ is the acoustic phonon frequency with
$v_{ph}$ being the phonon velocity,  and
$N_q$  the phonon 
occupation number
$N_q = {1}/({\exp(\beta \omega_{\vq}) -1})$.
The first (second) term is Eq.~(\ref{wq}) corresponds to the
absorption (emission) of an acoustic phonon of wave vector $\vq = \vk-\vk'$.
The matrix element $C(\vq)$ for the deformation potential is given by
\begin{equation}
|C(\vq)|^2 = \frac{D^2\hbar q}{2A\rho_m v_{ph}}\left [ 1- \left (
    \frac{q}{2k} \right )^2 \right ],
\end{equation}
where $D$ is the deformation potential coupling constant, $\rho_m$
is the graphene mass density, and $A$ is the area of the sample.
The scattering time due to the surface optical phonons is given by
\begin{eqnarray}
\frac{1}{\tau_{sp}(\varepsilon)} & = & \sum_{l\vk'}
 M_{0l}^2(\vk-\vk') g(\theta_{\vk,\vk'}) 
 [ (N_0 + f(\varepsilon))\delta(\varepsilon + \omega_{SOl} -
  \varepsilon_{\vk'}) \nonumber \\
& +& (N_0 +1 
      - f(\varepsilon))\delta(\varepsilon  -\omega_{SOl}-\varepsilon_{\vk'})  ]
\label{eq:2}
\end{eqnarray}
where the matrix elements of a surface optical phonon is given by
\begin{equation}
[M_{0l}(q)]^2 = \frac{2\pi e^2}{q \epsilon(q) }e^{-2qd}\frac{\omega_{SOl}}{2}\left [
  \frac{1}{\epsilon_{\infty}+1} - \frac{1}{\epsilon_0+1} \right ],
\end{equation}
where $d$ is the separation distance between graphene layer and
substrate, $\epsilon_0$
($\epsilon_{\infty}$) is the static (high frequency) dielectric
constant, and $\omega_{SOl}$ is the $l$-th kind of surface optical
phonon frequency.
Then by averaging over energy we have the conductivity as 
\begin{equation}
\sigma(T)=\frac{e^2}{2}\int d\varepsilon D(\varepsilon) v_F^2
\tau_{tot}(\varepsilon,T) \left (-\frac{\partial f(\varepsilon)}{\partial
  \varepsilon} \right ),
\end{equation}
where $D(\varepsilon)=2\varepsilon/\pi (\hbar v_F)^2$ is the density of
states of graphene.

\begin{figure}
\includegraphics[width=\columnwidth]{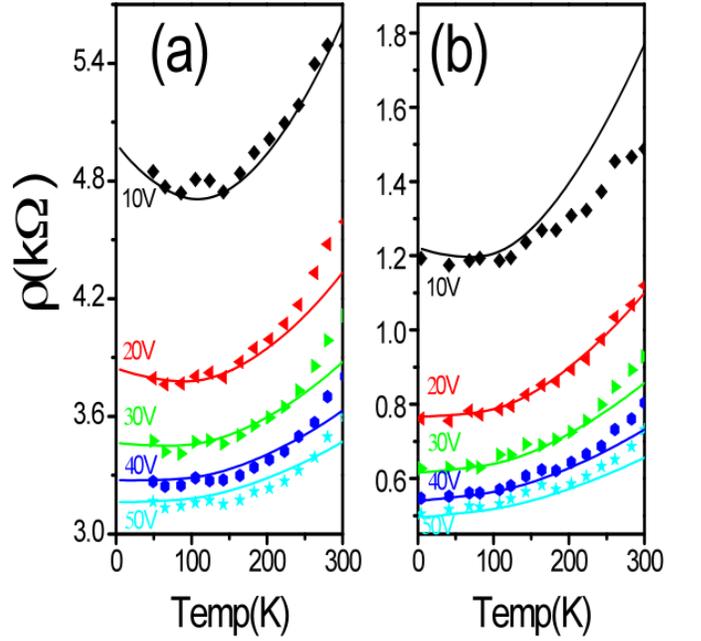}
\caption{
(color online) $\rho(T)$ at varied $\Delta V_g$. Scatters are experimental data and
  lines are calculation for (a) device A, (b) device B.  
}
\end{figure}

We find that the screened charge impurity scattering dominates over
phonon scattering at low T and strongly contributes to $\rho(T)$ for all
T. The acoustic phonon scattering gives rise to a linear T dependent
resistivity and its contribution to $\rho(T)$ is a quantitatively small
even at room temperature. It is only important for high mobility and
high density samples because it is independent of $n_g$ and sample
quality. It is known that the carriers in graphene are strongly
coupled to the surface optical phonons of a polar substrate such as
SiO$_2$. Consequently, the surface phonon scattering gives a large
contribution to $\rho(T)$ when the carrier temperature is high enough to
absorb or emit phonons. We find that the surface phonon scattering
plays an important role for $T > 200$K and the scattering rate strongly
depends on the distance between the 2D graphene layer and the
substrate.
 
We compare experimental and theoretical $\rho(T)$ in Fig. 3 incorporating
all effects we discussed above. We use the charged impurity densities
($n_i$) and the standard deviation of potential fluctuation ($s$), (a) $n_i =
1.8 \times 10^{12} cm^{−2}$ and $s=70$ meV, (b) $n_i = 4.5 \times
10^{11} cm^{−2}$ and $s=110$ meV. To get the charge impurity densities
we first fit our data at low temperature and high density limits
because in these limits all phonon scattering effects are exponetionally
suppressed and scattering by charged impurity dominates.
For acoustic phonon scattering we
use the phonon velocity $v_{ph}=2\times 10^6$ cm/s and 
the deformation potential coupling constant $D=19$ eV.\cite{hwang:2008} For surface
optical phonon scattering we use the two optical phonons in our
calculation with  $\omega_{sp1}=63$ meV and and $\omega_{sp2}=116$
meV.\cite{zou:2010} The overall resistivity scale depends on the 
impurity density or mobility value, but the qualitative trends in $\rho(T)$
arise from the basic aspects of the underlying scattering
mechanisms. The rise in $\rho$ with increasing T at high temperatures is a
direct result of the thermal weakening of screening and phonon
effects. The decrease in $\rho$ with temperature at low temperatures arises
from thermal activation in the presence of electron-hole puddles\cite{hwang:2010}.
The non-monotonicity of $\rho(T)$ at the intermediate temperatures arises
from the competition among activation, temperature dependent screening
and phonon scattering. The activated insulating behavior competes with
the metallic behavior induced by the temperature dependent screening
and phonon effects. When the potential fluctuation is smaller than the
chemical potential the activation behavior is suppressed. As a result
the total conductivity manifests a metallic behavior. However, for
large potential fluctuations the temperature dependence by activation
overwhelms the temperature dependence due to all other mechanisms and
the system shows insulating behavior. In our calculations we neglect
Anderson localization. But we cannot rule out that in the lowest
mobility sample localization may play a role in the insulating
behavior. Detailed quantitative understanding in this regime would
require a more sophisticated theory which includes higher order
electron-electron interactions and disorder induced localization
corrections, but our theoretical results
\cite{hwang:2010,hwang:2007,hwang:2009,hwang:2008,fratini:2008}  give an
excellent qualitative description of the data. We also measure the
resistivity (not shown) for very low mobility sample ($\mu=300$ $cm^2$/Vs)
showing a stronger insulating behavior at low temperatures than high
mobility samples of Fig. 3(a) and (b).

\section{DISCUSSION AND CONCLUSION}

In conclusion, temperature dependent transport of CVD grown graphene
is investigated. For low mobility samples, an anomalous insulating
behavior at low temperatures and non-monotonic at higher temperatures
are observed, in contrast to the monotonic temperature dependent
transport of high mobility exfoliation graphene. However, we note that
a strong insulating behavior can be obsered at low mobility exfoliated
graphene. Most important factor in observing an insulating temperature
dependent transport is the strength of potential fluctuations 
which are proportional to the charged impurity density
or inversely proportional to the mobility. If the sample mobility is
low enough the insulating behavior is notable for even exfoliation
graphene samples\cite{zhu:2009}. These are explained based on the transport
theory incorporating thermally activated transport of inhomogeneous
electron-hole puddles. Our results imply that carrier transport in CVD
graphene devices, which are vulnerable to the impurities or charged
defects during the transfer process, is strongly affected by
underlying environments, which gives rise to lower mobility in CVD
graphene than the exfoliated graphene since both CVD and exfoliated
graphene use the same substrate. To get higher mobility CVD graphene a
careful transfer process is required in order to keep the background
impurity density low.

We have shown, through experimental measurements and detailed
comparison with theoretical calculations, that the temperature
dependence of graphene resistivity can be understood as arising from a
complex interplay among the physical mechanisms of phonon scattering,
screening, Fermi surface averaging, and activation across potential
barriers in inhomogeneous puddles.  The first (last) two mechanisms
lead to metallic (insulating) temperature dependence with resistivity
increasing (decreasing) with temperature, and therefore, the net
effect could be either an insulating or a metallic temperature
dependence or a non-monotonic combination of both, depending on the
details of sample mobility, density, and the temperature regime.  With
increasing (decreasing) density and mobility, puddle effects are
suppressed (enhanced), leading to stronger metallicity.  With
increasing temperature, phonon effects eventually dominate, so at the
highest temperature (and not too low densities), the system always
manifests metallic behavior with increasing resistivity with
increasing temperature.  At the Dirac point, activation in puddles
dominate and the insulating behavior is generic.  For intermediate
temperatures and densities, the generic behavior of resistivity as a
function of temperature is non-monotonic, as predicted in
ref.~\onlinecite{hwang:2010}, with
the resistivity decreasing with increasing temperature at low
temperatures and then eventually increasing with temperature (except
at the lowest densities where puddles are dominant even at room
temperatures) at higher temperatures.  The crossover temperature for
the transition from insulating to metallic transport behavior
increases with decreasing mobility and/or carrier density.  All of
these observed effects are completely consistent with the predictions
of the theory developed recently\cite{hwang:2010,li:2011}.

\section*{Acknowledgment}

The work at the University of Maryland is supported by NRI-SWAN and US-ONR.

\end{document}